\pdfoutput=1
\documentclass[11pt]{article}

\usepackage{authblk}
\usepackage{emnlp2021}

\usepackage{times}
\usepackage{latexsym}
\usepackage{caption}
\usepackage{subcaption}
\usepackage{url}

\usepackage[T1]{fontenc}
\usepackage[utf8]{inputenc}

\usepackage{microtype}

\usepackage{amstext} % for \text macro
\usepackage{array}   % for \newcolumntype macro
\usepackage{bbding}
\usepackage{graphicx}
\title{DS4DH at TREC Health Misinformation 2021: Multi-Dimensional Ranking Models with Transfer Learning and Rank Fusion}

\author[1]{Boya Zhang} 
\author[2,3]{Nona Naderi}
\author[1,2]{Fernando Jaume-Santero}
\author[1,2,3]{Douglas Teodoro}
\affil[1]{Department of Radiology and Medical Informatics, University of Geneva \authorcr { \tt <first>.<last>@unige.ch}}
\affil[2]{HES-SO University of Applied Sciences and Arts of Western Switzerland \authorcr { \tt <first>.<last>@hesge.ch}}
\affil[3]{Swiss Institute of Bioinformatics}

\begin{document}
\maketitle

\begin{abstract}
This paper describes the work of the Data Science for Digital Health (DS4DH) group at the TREC Health Misinformation Track 2021. The TREC Health Misinformation track focused on the development of retrieval methods that provide relevant, correct and credible information for health related searches on the Web. In our methodology, we used a two-step ranking approach that includes \textit{i)} a standard retrieval phase, based on  BM25 model, and \textit{ii)} a re-ranking phase, with a pipeline of models focused on the usefulness, supportiveness and credibility dimensions of the retrieved documents. To estimate the usefulness, we classified the initial rank list using pre-trained language models based on the transformers architecture fine-tuned on the MS MARCO corpus. To assess the supportiveness, we utilized BERT-based models fine-tuned on scientific and Wikipedia corpora. Finally, to evaluate the credibility of the documents, we employed a random forest model trained on the Microsoft Credibility dataset combined with a list of credible sites. The resulting ranked lists were then combined using the Reciprocal Rank Fusion algorithm to obtain the final list of useful, supporting and credible documents. Our approach achieved competitive results, being top-2 in the compatibility measurement for the automatic runs. Our findings suggest that integrating automatic ranking models created for each information quality dimension with transfer learning can increase the effectiveness of health-related information retrieval.

\end{abstract}

\section{Introduction} 
\label{intro}
The purpose of the TREC Health Misinformation Track\footnote{\url{https://trec-health-misinfo.github.io/}} is to develop retrieval systems that provide relevant and correct information for health-related Web searches. The challenge provides 50 health-related topics, among which only 35 topics are evaluated. As illustrated in Table \ref{anothertable}, each topic contains a query, a description, a narrative, a stance and an evidence. For automatic runs, only the query and the description are used. The TREC Health Misinformation corpus contains one billion English documents extracted from the April 2019 snapshot of Common Crawl \cite{2020t5}.

For a document to be relevant and correct, the TREC Health Misinformation challenge considers three levels of information quality: (1) \textit{usefulness}, that is, whether a document contains relevant information to answer a topic's question; (2) \textit{supportiveness}, that is, whether a document contains supportive information for the descriptions marked as helpful or dissuasive information for the descriptions marked as unhelpful; and (3) \textit{credibility}, that is, whether an information source document is considered credible in the field of knowledge. 

This paper describes submissions of DS4DH to the TREC Health Misinformation 2021, which achieved top-2 performance in the compatibility assessment for the automatic runs. Section \ref{met} introduces our methodology based on a two-step approach of document retrieval for the multi-dimensional retrieval task and the specific evaluation criteria used to assess these dimensions with a single ranked list. Section \ref{res} presents and discusses our results. Finally, Section \ref{con} concludes this paper and proposes future studies.

\begin{table*}[ht!]
\centering
\renewcommand{\arraystretch}{1}
\scalebox{0.8}{\begin{tabular}{c | p{9.5cm}}
\hline
\multicolumn{2}{c}{Topic 101} \\
\hline
Query & ankle brace achilles tendonitis \\
\hline
Description & Will wearing an ankle brace help heal achilles tendonitis?\\
\hline
Narrative & Achilles tendonitis is a condition where one experiences pain in the Achilles tendon located near the heel. An ankle brace is usually worn around the ankles to protect and limit movement. A very useful document would discuss the effectiveness of using ankle braces to help heal Achilles tendonitis. A useful document would help a user make a decision about the use of ankle braces for treating tendonitis by providing information about recommended treatments for Achilles tendonitis, ankle braces, or both.\\
\hline
Stance & unhelpful \\
\hline
Evidence & https://www.ncbi.nlm.nih.gov/pmc/articles/PMC3134723/ \\
\hline
\end{tabular}}
\caption{Topic Example.}
\label{anothertable}
\end{table*}

\section{Methods} \label{met}

\begin{figure*}[!tp]
    %\centering
    \includegraphics[scale=0.14]{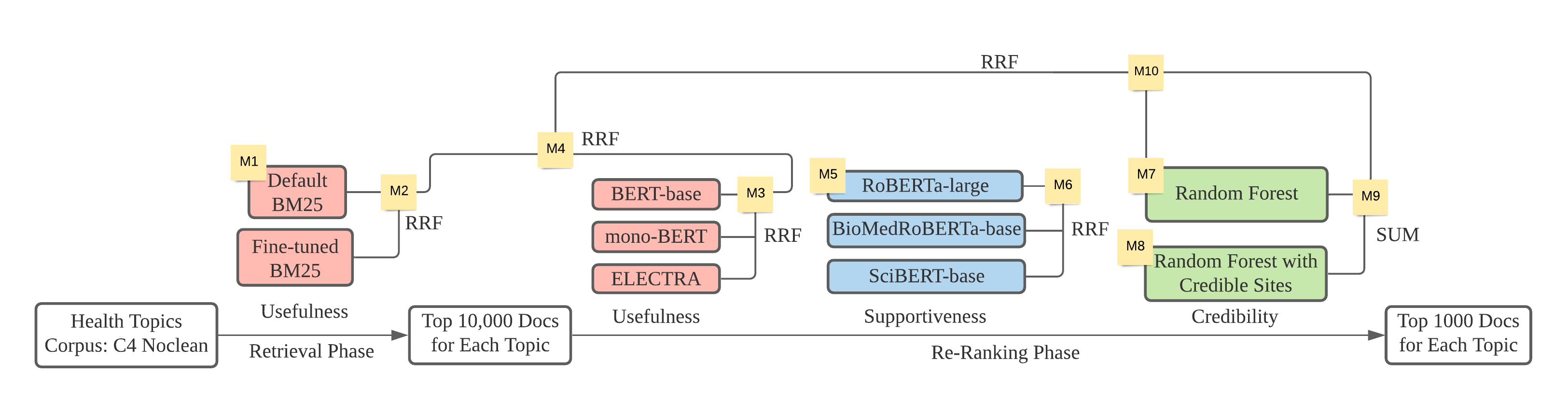}
    \caption{Retrieval Pipeline. M1: BM25 baseline. M2: M1 combined with a fine-tuned index using a silver standard query relevance. M3: a re-ranked list using masked language models (MLMs) fine-tuned on the MS MARCO dataset. M4: a combination BM25 and MLMs. M5: a RoBERTa model fine-tuned on the FEVER+SciFact fact-checking datasets. M6: A combination of three MLMs trained on the FEVER+SciFact fact-checking corpora. M7: a random forest model trained on the Microsoft Credibility dataset to predict a site's credibility. M8: a list of credible websites scrapped from the Health-on-Net search engine for the challenge's queries.  M9: a linear combination of M7 and M8 models. M10: a combination of model M4, M6 and M9. All ranking combinations, apart from M9, were created using the reciprocal ranking fusion (RRF) algorithm with the k parameter set to 60.}
    \label{runsoverview}
\end{figure*}

The overview of our pipeline is presented in Figure~\ref{runsoverview}. 
First, in the retrieval phase, we extracted 10,000 documents using a BM25 model \cite{robertson2009probabilistic}. Second, in the re-ranking phase, we estimated relevance of the documents according to the usefulness, supportiveness and credibility scores. At the end, 7 runs with different model combinations were submitted. 

\subsection{Retrieval Phase}

We extracted 10,000 documents using a BM25 model with standard \cite{robertson2009probabilistic} and fine-tuned parameters. For the fine-tuning version, as the query relevance was not available, we created topics, that is, query + description, for a set of indexed documents using the a keyword2query and doc2query approaches proposed by Bennani-Smires \textit{et al.} \cite{bennani2018simple} and Nogueira \textit{et al.} \cite{nogueira2019document}, respectively. Then, a known-item search approach was applied using the silver topics and the BM25 parameters was fine-tuned using a grid search. This resulted in two initial ranking lists.

\subsection{Re-Ranking Phase}

For re-ranking the retrieved documents, we use a set of machine learning models trained to classify documents according to the usefulness, supportiveness, and credibility criteria.

\paragraph{Usefulness} 
To improve the usefulness dimension of the retrieved documents, we implemented re-ranking models based pre-trained language models fine-tuned on the MS MARCO dataset: BERT-base \cite{li2021parade}, mono-BERT-large \cite{nogueira2019multistage} and ELECTRA \cite{clark2020electra}. While BM25 provides a strong baseline for usefulness, it does not consider the relation and context of words. Thus, the pre-trained language models are used to enhance the quality of the original ranking \cite{teodoro2021information} as similarly shown to improve other natural language processing tasks like named entity recognition \cite{naderi2021ensemble}. Given a topic and a document, the language model infers whether the document is relevant or not to the topic.

\paragraph{Supportiveness} In this information quality dimension, documents are identified under three levels: 1) \textit{supportive} - the document supports the treatment; 2) \textit{dissuades} - the document refutes the treatment; 3) \textit{neutral} - the document does not contain enough information to make the decision. We want documents that are either  supportive or dissuasive on the top of the ranking list, which means that correct or factual documents are boosted and misinforming documents should be downgraded.

\begin{figure*}[!tp]
    \centering
    \includegraphics[scale=0.75]{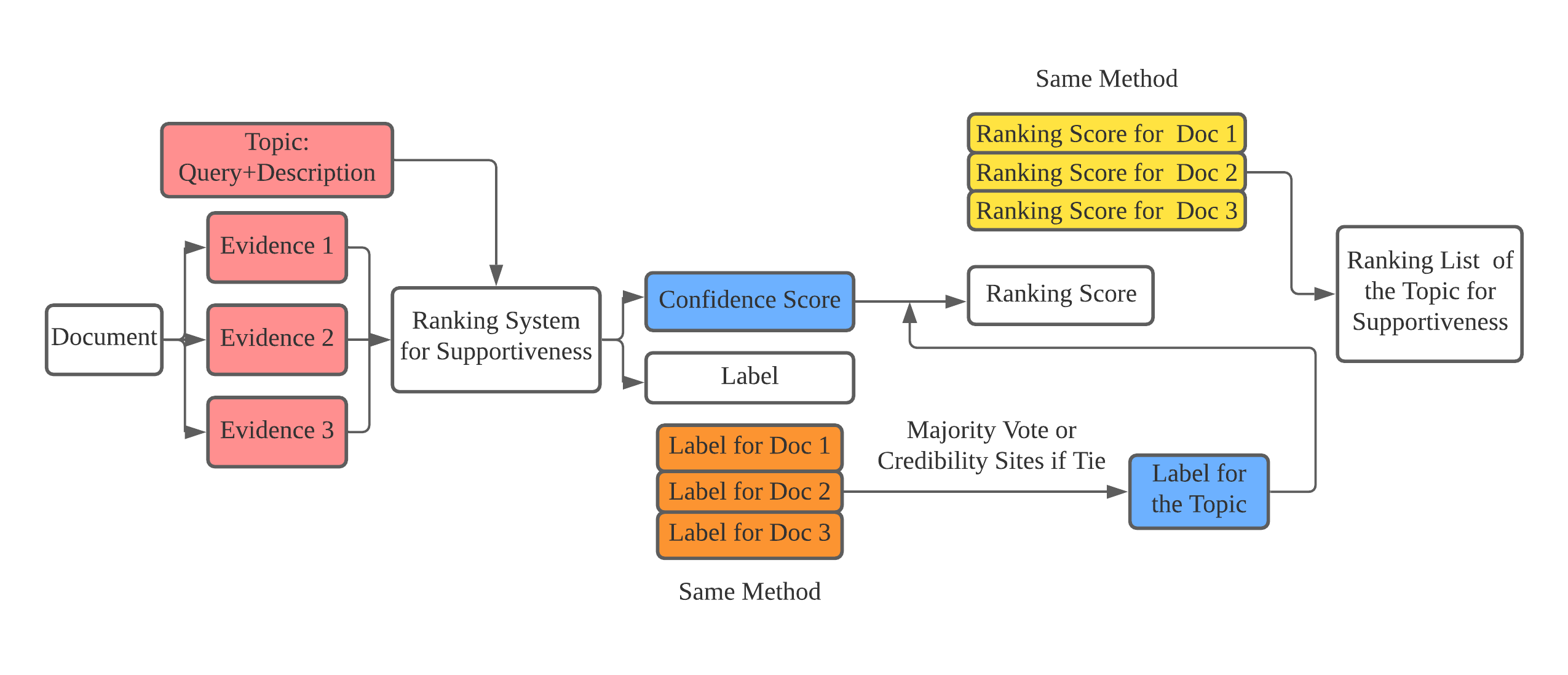}
    \caption{Re-Ranking Pipeline for Supportiveness. When the supportive and dissuasive documents are for one topic, we use the majority vote to decide the correct label. The document ranking score of the incorrect side becomes negative. We bring in the credibility sites when tie.}
    \label{afigurename}
\end{figure*}

The supportiveness dimension shares similarities with claim-checking models, which take a claim and a document as the information source, and \textit{validate} or \textit{refute} the claim based on document content \cite{stammbach2021choice}. Their main difference is that for claim-checking models, we assume that the information source is always correct. Thus, for the supportiveness criterion, we add a further classification step, which evaluates the documents as correct or incorrect. To do so, we used a k-nearest neighbors algorithm \cite{teodoro2010automatic} based on the top-k assignments provided by the claim-checking models, that is, a majority vote is used to decide whether the treatment should be supported or dissuaded. Then, higher rank is given to the correct supportive/dissuasive documents, medium rank is given to the neutral documents and lower rank is given to the incorrect supportive/dissuasive documents. The details are shown in Figure \ref{afigurename} and Table \ref{aname}.

We used three models from the Scientific Claim Verification task \cite{wadden2020fact} to classify the treatments: RoBERTa-Large \cite{liu2019roberta}, BioMedRoBERTa-base \cite{domains} and SciBERT-base \cite{Beltagy2019SciBERT}. These models were trained on either scientific or large English corpora and fine-tuned on the FEVER \cite{thorne2018fever} and SciFact \cite{wadden2020fact} datasets. The information that these models learned from previous corpora benefits our ranking task through transfer learning.

\begin{table*}[ht!]
\centering
\renewcommand{\arraystretch}{1}
\scalebox{0.8}{\begin{tabular}{c c c|c c c}
    \hline
        \multicolumn{3}{c|}{Evidences 
            ($E_{1}, ... , E_{i}$)} & \multicolumn{3}{c}{Document
            ($D_{j}$)} \\ \hline
        Supports & Dissuades & Neutral & Label & Confidence Score &  Ranking Score \\   \hline
        \Checkmark & \XSolidBrush & ~ & Supports & $S_{max}$ &  $S_{max}$ or -$S_{max}$ \\ 
        \XSolidBrush & \Checkmark & ~ & Dissuades & $S_{max}$ &  $S_{max}$ or -$S_{max}$ \\ 
        \XSolidBrush & \XSolidBrush & \Checkmark & ~ & $1-\frac{1}{N}\sum_{i=1}^{N}S_{i}$ &  $1-\frac{1}{N}\sum_{i=1}^{N}S_{i}$ \\ 
        \XSolidBrush & \XSolidBrush & \XSolidBrush & ~ & -3 &  -3 \\ 
        \Checkmark & \Checkmark & ~ & ~ & -2 &  -2 \\ \hline
    \end{tabular}}

\caption{Ranking Score for Supportiveness. When the document only includes neutral evidence, the document confidence score is $1-\frac{1}{N}\sum_{i=1}^{N}S_{i}$, where $S_1, S_2, ..., S_i$ is the confidence score of each evidence in the document. When the document contains both supports and dissuades evidences without considering the neutral cases, the document confidence score is -2. When evidences are either support or dissuades without considering the neutral cases, the document confidence score is $S_{max}$, where $S_{max}$ is highest among the confidence score of each evidence in the document. When the qualified evidence is void, the document confidence score is -3.}
\label{aname}
\end{table*}

\paragraph{Credibility}
For estimating credibility, we develop a random forest classifier trained on the Microsoft Credibility dataset \cite{schwarz2011augmenting} with a set of features, such as readability, openpage rank\footnote{https://www.domcop.com/openpagerank/documentation} and number of CSS style sheets. The dataset consists of 1,000 Web pages on five topics of Health, Politics, Finance, Environmental Science, and Celebrity News. The Web pages are manually rated with credibility scores between 1 ("very non-credible") and 5 ("very credible"). \footnote{A credibile Web page is defined as "a page whose information one can accept as the truth without needing to look elsewhere."} We convert these scores for a binary classification setting -- that is, the scores of 4 and 5 are considered as 1 or \textit{credible} and scores of 1, 2, and 3 are considered as 0 or \textit{non-credible}.   
For the readability score, we rely on SMOG index, which estimates the years of education an average person needs to understand a piece of writing. Following \cite{schwarz2011augmenting}, we retrieve a Web page's PageRank and use it as a feature to train the classifier.  We further use the number of CSS style definitions for estimating the effort for the design of a Web page (\citet{olteanu2013web} showed the effectiveness of this feature). Furthermore, a
list of credible websites scrapped from the Health On the Net search engine~\footnote{\url{www.hon.ch}} for the challenge’s queries is combined with the baseline model to explore better performance. The result of the classifier was added with a unitary value for the Health on the Net credible sites.

\subsection{Submitted Runs}

\paragraph{Run 1:} Baseline run. A combination of M1, M5 and M7 in Figure \ref{runsoverview}. This automatic run was created using a model based on Reciprocal Rank Fusion (RRF) \cite{cormack2009reciprocal} of three models: i) usefulness, created using a default BM25, ii) supportiveness, created using a RoBERTa large model fine-tuned on the FEVER and SciFact corpus, and iii) credibility, created using a credibility random forest classifier as described in Section\ref{met}.

\paragraph{Run 2:} A combination of M2, M6 and M9 in Figure \ref{runsoverview}. This automatic run was created using a rank fusion based on RRF of three models: i) usefulness, created using a combined default BM25 with a fine-tuned BM25 model using known item search with query and description been generated using transfer learning from language models, ii) supportiveness, created using a combined rank of three transformer-based models fine-tuned on the FEVER and SciFact corpus, and iii) credibility, created using a random forest classifier trained on the Microsoft Credibility dataset combined with a list of credible sites.

\paragraph{Run 3:} A combination of M3, M6 and M9 in Figure \ref{runsoverview}. This automatic run was created using a rank fusion based on RRF of three models: i) usefulness, created using a combination of three transformed-based language models trained on the MS MARCO corpus\footnote{\url{https://microsoft.github.io/msmarco/}}, ii) supportiveness, created using a combined rank of three transformer-based models fine-tuned on the FEVER and SciFact corpus, and iii) credibility, created using a random forest model trained on the Microsoft Credibility dataset combined with a list of credible sites.

\paragraph{Run 4:} A combination of M4, M5 and M9 in Figure \ref{runsoverview}. This automatic run was created using a rank fusion based on RRF of three models: i) usefulness, created using a combined BoW model with three transformed-based language models trained on the MS MARCO corpus, ii) supportiveness, created using a RoBERTa large model fine-tuned on the FEVER and SciFact corpus, and iii) credibility, created using a random forest model trained on the Microsoft Credibility dataset combined with a list of credible sites.

\paragraph{Run 5:} A combination of M4, M6 and M7 in Figure \ref{runsoverview}. This automatic run was created using a rank fusion based on RRF of three models: i) usefulness, created using a combined BoW model with three transformed-based language models trained on the MS MARCO corpus, ii) supportiveness, created using a combined rank of three transformer-based models fine-tuned on the FEVER and SciFact corpus, and iii) credibility, create using a random forest model trained on the Microsoft Credibility dataset.

\paragraph{Run 6:} A combination of M4, M6 and M9 in Figure \ref{runsoverview}. This automatic run was created using a rank fusion based on RRF of three models: i) usefulness, created using a combined BoW model with three transformed-based language models trained on the MS MARCO corpus, ii) supportiveness, created using a combined rank of three transformer-based models fine-tuned on the FEVER and SciFact corpus, and iii) credibility, created using a random forest classifier combined with a list of credible sites.

\paragraph{Run 7:} A combination of all the individual models in Figure \ref{runsoverview}. This automatic run was created using a rank fusion based on RRF of the individual models used to create the i) usefulness (5 individual models), ii) supportiveness (3 individual models), and iii) credibility (2 individual models). 

\subsection{Evaluation Methods}

The assessments of this year's TREC Health Misinformation were divided into ``compatibility with helpful'' (help) and ``compatibility with harmful'' (harm) metrics. In order to evaluate the system's ability to have higher levels for helpful information and lower levels for harmful information, the ``compatibility with harmful'' results were subtracted from the ``compatibility with helpful'' results, which is marked as ``help-harm''. This orders systems correctly with similar helpful compatibility and lower harmful compatibility. For more details, please see \cite{clarke2021assessing}.

\section{Results and Discussion} \label{res}

\begin{figure}[tp!]
    \centering
    \includegraphics[width=0.45\textwidth]{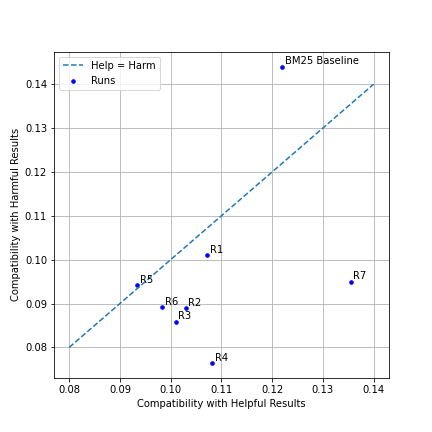}
    \caption{Relations between `Help' and `Harm' for Submitted Runs.}
    \label{rchch}
\end{figure}

The official challenge results are shown in Table \ref{acds} and Figure \ref{rchch}. The `Best Help', `Best Harm' and `Best Hp-Hm' are are the highest ``compatibility with helpful'' (help), lowest ``compatibility with harmful'' (harm) and highest ``help-harm'' from the list of top-3 automatic runs from each group. Note that `Run 1' to `Run 7' are runs submitted by the DS4DH group. `BM25 Baseline' is the baseline model by the TREC Health Misinformation. 

In our submitted runs, the best harm is in `Run 4', the best help and help-harm are both in `Run 7'. This is as expected since Run 7 combines all the individual models to maximize the system ranking ability. The helpful compatibility is 0.136 which is comparable to the `Best Help', and the harmful compatibility 0.095 is significantly low, which is similar to the `Best Harm'. Therefore, we obtained help-harm at 0.041, which is right after the overall `Best Hp-Hm', together with superior performance on lower ranking harmful information.

The `Best Help' run achieved in the challenge gives the helpful compatibility at 0.203. However, for this run, the harmful compatibility is also as high as 0.168. Therefore, the help-harm is 0.034, which can distinguish the helpful and harmful information but needs improvement by lower ranking the harmful one. The `Best Harm' gives the harmful compatibility at 0.022, which is significantly lower than other runs. However, the helpful compatibility for this run is only 0.006, resulting in a help-harm of -0.016, in a sense that neither helpful nor harmful information is retrieved. On the other hand, the `Best Hp-Hm' gives relatively higher helpful compatibility at 0.195 and relatively lower compatibility at 0.153. In the end, the best help-harm 0.043 is obtained.

\begin{table}[tp!]
\centering
\renewcommand{\arraystretch}{1}
\scalebox{0.8}{\begin{tabular}{c c c c}
\hline
Run & Help & Harm & Help-Harm \\ 
\hline
Best Help & \textbf{0.203} & 0.168 & 0.034\\ 
Best Harm & 0.006 &	\textbf{0.022} & -0.016 \\
Best Hp-Hm & 0.195 & 0.153 & \textbf{0.043} \\
\hline
Run 1 & 0.107 & 0.101 & 0.006 \\
Run 2 & 0.103 & 0.089 & 0.014 \\
Run 3 & 0.101 & 0.086 & 0.015 \\ 
Run 4 & 0.108 & \textbf{0.076} & 0.032 \\ 
Run 5 & 0.093 & 0.094 & -0.001 \\ 
Run 6 & 0.098 & 0.089 & 0.009 \\ 
Run 7 & \textbf{0.136} & 0.095 & \textbf{0.041} \\
\hline
BM25 Baseline & 0.122 & 0.144 & -0.022 \\
\hline
\end{tabular}}
\caption{Averaged Compatibility of Submitted Runs.}
\label{acds}
\end{table}

\subsection{Models Analysis}

\begin{figure}[tp!]
    \centering
    \includegraphics[width=0.45\textwidth]{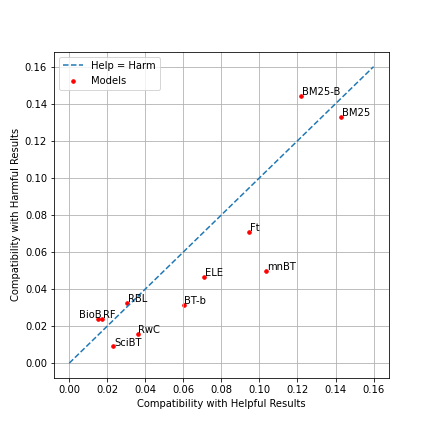}
    \caption{Relations between `Help' and `Harm' for Individual Models. ELE: ELECTRA. RwC: Random Forest with Credibility Sites. mnBT: monoBERT. RF: Random Forest. BioB: BioMedRoBERTa-base. RBL: RoBERTa-Large. SciBT: SciBERT. BM25: Standard BM25. Ft: Fine-tuned BM25. BM25-B: BM25 Baseline by TREC Health Misinformation}
    \label{fig:eval-comp}
\end{figure}

To understand the performance of each component, we study the compatibility of individual models. The results are depicted in Table \ref{acim} and Figure \ref{fig:eval-comp}. The BM25 model for usefulness in the retrieval phase gives highest helpful compatibility with 0.143. This is beneficial since at this phase, overall compatibility is more critical. The monoBERT model for usefulness in the re-ranking phase gives highest help-harm compatibility at 0.053, which is the top-1 result compared to the automatic runs submitted by the participants. This indicates that the model can effectively differentiate between helpful and harmful information.
The SciBERT-base model for supportiveness gives lowest harmful compatibility at 0.009 with satisfactory helpful compatibility. This demonstrates the model's ability to identify misinformation.

As a result of the foregoing analysis, further experiments could be explored: 1) in the document retrieval phase, BM25 should be the prior model. 2) in the re-rankning phase: prioritising the monoBERT for usefulness, the SciBERT-base model for supportiveness, the 'Random Forest with Credibility Sites' model for credibility.

\begin{table*}[tp!]
\centering
\renewcommand{\arraystretch}{1}
\scalebox{0.8}{\begin{tabular}{l|l|c c c}
    \hline
         ~ & Individual Model & Help & Harm & Help-Harm \\ \hline
         Usefulness & BM25 & \textbf{0.143} & 0.133 & 0.010 \\
         (Retrieval) & Fine-tuned BM25 & 0.095 & 0.071 & 0.024 \\ \hline
        Usefulness  & BERT-base & 0.060 & 0.031 & 0.029 \\ 
         (Re-Ranking)& monoBERT & 0.103 & 0.050 & \textbf{0.053} \\
         ~ & ELECTRA & 0.071 & 0.047 & 0.024 \\ \hline
         Supportiveness & BioMedRoBERTa-base & 0.015 & 0.024 & -0.009 \\
         ~ & RoBERTa-large & 0.031 & 0.033 & -0.002 \\
         ~ & SciBERT-base & 0.023 & \textbf{0.009} & 0.014 \\ \hline
         Credibility & RandomForest & 0.017 & 0.024 & -0.007 \\
         ~ & RFwithCredibilitySites & 0.036 & 0.016 & 0.020 \\ \hline
        TREC HM & BM25 Baseline & 0.122 & 0.144 & -0.022 \\ \hline
    \end{tabular}}
\caption{Averaged Compatibility of Individual Models.}
\label{acim}
\end{table*}

\section{Conclusion} \label{con}

In the average compatibility measurement of automatic runs, our contributions came out on top-2. Our findings imply that combining automatic ranking models for each information quality dimension with transfer learning can improve the quality of health-related information retrieval by allowing the proper documents to be retrieved while discarding the incorrect ones. The RRF algorithm is a robust alternative for combining ranks when no trained set is available. Further empirical approaches could be to fine-tune our models based on topic content and generate manual runs and to re-rank the top 10\% of the retrieved documents to reduce the possibility of bringing the harmful documents from the bottom of the ranking list to the top.

\bibliography{emnlp2021}

\begin{thebibliography}{20}
\expandafter\ifx\csname natexlab\endcsname\relax\def\natexlab#1{#1}\fi

\bibitem[{Beltagy et~al.(2019)Beltagy, Lo, and Cohan}]{Beltagy2019SciBERT}
Iz~Beltagy, Kyle Lo, and Arman Cohan. 2019.
\newblock Scibert: A pretrained language model for scientific text.
\newblock In \emph{Proceedings of the 2019 Conference on Empirical Methods in
  Natural Language Processing and the 9th International Joint Conference on
  Natural Language Processing (EMNLP-IJCNLP)}, pages 3615--3620.

\bibitem[{Bennani-Smires et~al.(2018)Bennani-Smires, Musat, Hossmann,
  Baeriswyl, and Jaggi}]{bennani2018simple}
Kamil Bennani-Smires, Claudiu Musat, Andreea Hossmann, Michael Baeriswyl, and
  Martin Jaggi. 2018.
\newblock Simple unsupervised keyphrase extraction using sentence embeddings.
\newblock In \emph{Proceedings of the 22nd Conference on Computational Natural
  Language Learning}, pages 221--229.

\bibitem[{Clark et~al.(2020)Clark, Luong, Le, and Manning}]{clark2020electra}
Kevin Clark, Minh-Thang Luong, Quoc~V. Le, and Christopher~D. Manning. 2020.
\newblock Electra: Pre-training text encoders as discriminators rather than
  generators.
\newblock \emph{arXiv preprint arXiv:2003.10555}.

\bibitem[{Clarke et~al.(2021)Clarke, Vtyurina, and
  Smucker}]{clarke2021assessing}
Charles~LA Clarke, Alexandra Vtyurina, and Mark~D Smucker. 2021.
\newblock Assessing top-preferences.
\newblock \emph{ACM Transactions on Information Systems (TOIS)}, 39(3):1--21.

\bibitem[{Cormack et~al.(2009)Cormack, Clarke, and
  Buettcher}]{cormack2009reciprocal}
Gordon~V Cormack, Charles~LA Clarke, and Stefan Buettcher. 2009.
\newblock Reciprocal rank fusion outperforms condorcet and individual rank
  learning methods.
\newblock In \emph{Proceedings of the 32nd international ACM SIGIR conference
  on Research and development in information retrieval}, pages 758--759.

\bibitem[{Gururangan et~al.(2020)Gururangan, Marasović, Swayamdipta, Lo,
  Beltagy, Downey, and Smith}]{domains}
Suchin Gururangan, Ana Marasović, Swabha Swayamdipta, Kyle Lo, Iz~Beltagy,
  Doug Downey, and Noah~A. Smith. 2020.
\newblock Don't stop pretraining: Adapt language models to domains and tasks.
\newblock In \emph{Proceedings of ACL}.

\bibitem[{Li et~al.(2020)Li, Yates, MacAvaney, He, and Sun}]{li2021parade}
Canjia Li, Andrew Yates, Sean MacAvaney, Ben He, and Yingfei Sun. 2020.
\newblock Parade: Passage representation aggregation for document reranking.
\newblock \emph{arXiv preprint arXiv:2008.09093}.

\bibitem[{Liu et~al.(2019)Liu, Ott, Goyal, Du, Joshi, Chen, Levy, Lewis,
  Zettlemoyer, and Stoyanov}]{liu2019roberta}
Yinhan Liu, Myle Ott, Naman Goyal, Jingfei Du, Mandar Joshi, Danqi Chen, Omer
  Levy, Mike Lewis, Luke Zettlemoyer, and Veselin Stoyanov. 2019.
\newblock Roberta: A robustly optimized bert pretraining approach.
\newblock \emph{arXiv preprint arXiv:1907.11692}.

\bibitem[{Naderi et~al.(2021)Naderi, Knafou, Copara, Ruch, and
  Teodoro}]{naderi2021ensemble}
Nona Naderi, Julien Knafou, Jenny Copara, Patrick Ruch, and Douglas Teodoro.
  2021.
\newblock Ensemble of deep masked language models for effective named entity
  recognition in health and life science corpora.
\newblock \emph{Frontiers in Research Metrics and Analytics}, 6:689803--689803.

\bibitem[{Nogueira et~al.(2019{\natexlab{a}})Nogueira, Yang, Cho, and
  Lin}]{nogueira2019multistage}
Rodrigo Nogueira, Wei Yang, Kyunghyun Cho, and Jimmy Lin. 2019{\natexlab{a}}.
\newblock Multi-stage document ranking with bert.
\newblock \emph{arXiv e-prints}, pages arXiv--1910.

\bibitem[{Nogueira et~al.(2019{\natexlab{b}})Nogueira, Yang, Lin, and
  Cho}]{nogueira2019document}
Rodrigo Nogueira, Wei Yang, Jimmy Lin, and Kyunghyun Cho. 2019{\natexlab{b}}.
\newblock Document expansion by query prediction.
\newblock \emph{arXiv e-prints}, pages arXiv--1904.

\bibitem[{Olteanu et~al.(2013)Olteanu, Peshterliev, Liu, and
  Aberer}]{olteanu2013web}
Alexandra Olteanu, Stanislav Peshterliev, Xin Liu, and Karl Aberer. 2013.
\newblock Web credibility: Features exploration and credibility prediction.
\newblock In \emph{European conference on information retrieval}, pages
  557--568. Springer.

\bibitem[{Raffel et~al.(2020)Raffel, Shazeer, Roberts, Lee, Narang, Matena,
  Zhou, Li, and Liu}]{2020t5}
Colin Raffel, Noam Shazeer, Adam Roberts, Katherine Lee, Sharan Narang, Michael
  Matena, Yanqi Zhou, Wei Li, and Peter~J. Liu. 2020.
\newblock \href {http://jmlr.org/papers/v21/20-074.html} {Exploring the limits
  of transfer learning with a unified text-to-text transformer}.
\newblock \emph{Journal of Machine Learning Research}, 21(140):1--67.

\bibitem[{Robertson and Zaragoza(2009)}]{robertson2009probabilistic}
Stephen Robertson and Hugo Zaragoza. 2009.
\newblock \emph{The probabilistic relevance framework: BM25 and beyond}.
\newblock Now Publishers Inc.

\bibitem[{Schwarz and Morris(2011)}]{schwarz2011augmenting}
Julia Schwarz and Meredith Morris. 2011.
\newblock Augmenting web pages and search results to support credibility
  assessment.
\newblock In \emph{Proceedings of the SIGCHI conference on human factors in
  computing systems}, pages 1245--1254.

\bibitem[{Stammbach et~al.(2021)Stammbach, Zhang, and
  Ash}]{stammbach2021choice}
Dominik Stammbach, Boya Zhang, and Elliott Ash. 2021.
\newblock The choice of knowledge base in automated claim checking.
\newblock \emph{arXiv}, pages 2111--07795.

\bibitem[{Teodoro et~al.(2021)Teodoro, Ferdowsi, Borissov, Kashani, Alvarez,
  Copara, Gouareb, Naderi, Amini et~al.}]{teodoro2021information}
Douglas Teodoro, Sohrab Ferdowsi, Nikolay Borissov, Elham Kashani,
  David~Vicente Alvarez, Jenny Copara, Racha Gouareb, Nona Naderi, Poorya
  Amini, et~al. 2021.
\newblock Information retrieval in an infodemic: the case of covid-19
  publications.
\newblock \emph{Journal of medical Internet research}, 23(9):e30161.

\bibitem[{Teodoro et~al.(2010)Teodoro, Gobeill, Pasche, Ruch, Vishnyakova, and
  Lovis}]{teodoro2010automatic}
Douglas Teodoro, Julien Gobeill, Emilie Pasche, P~Ruch, D~Vishnyakova, and
  Christian Lovis. 2010.
\newblock Automatic ipc encoding and novelty tracking for effective patent
  mining.
\newblock In \emph{The 8th NTCIR Workshop Meeting on Evaluation of Information
  Access Technologies: Information Retrieval, Question Answering, and
  Cross-Lingual Information Access}.

\bibitem[{Thorne et~al.(2018)Thorne, Vlachos, Christodoulopoulos, and
  Mittal}]{thorne2018fever}
James Thorne, Andreas Vlachos, Christos Christodoulopoulos, and Arpit Mittal.
  2018.
\newblock Fever: a large-scale dataset for fact extraction and verification.
\newblock In \emph{Proceedings of the 2018 Conference of the North American
  Chapter of the Association for Computational Linguistics: Human Language
  Technologies, Volume 1 (Long Papers)}, pages 809--819.

\bibitem[{Wadden et~al.(2020)Wadden, Lin, Lo, Wang, van Zuylen, Cohan, and
  Hajishirzi}]{wadden2020fact}
David Wadden, Shanchuan Lin, Kyle Lo, Lucy~Lu Wang, Madeleine van Zuylen, Arman
  Cohan, and Hannaneh Hajishirzi. 2020.
\newblock Fact or fiction: Verifying scientific claims.
\newblock In \emph{Proceedings of the 2020 Conference on Empirical Methods in
  Natural Language Processing (EMNLP)}, pages 7534--7550.

\end{thebibliography}
\bibliographystyle{acl_natbib}

\appendix
\onecolumn

\setcounter{figure}{0} \renewcommand{\thefigure}{A.\arabic{figure}} \setcounter{table}{0} \renewcommand{\thetable}{A.\arabic{table}}

\section{Additional Illustration for Compatibility Measurement}

\begin{figure*}[!h]
    \centering
    \includegraphics[scale=0.5]{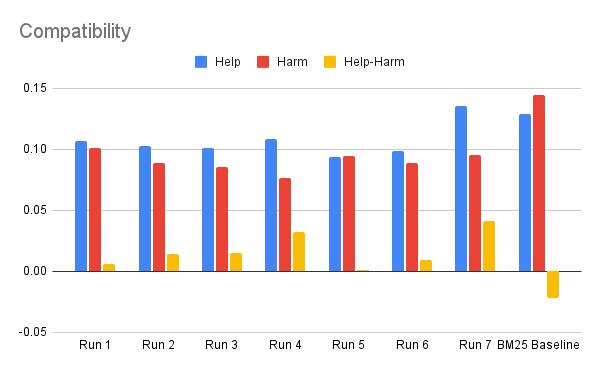}
    \caption{Runs Evaluated with Compatibility.}
    \label{rewc}
\end{figure*}

\begin{figure*}[!h]
    \centering
    \includegraphics[scale=0.42]{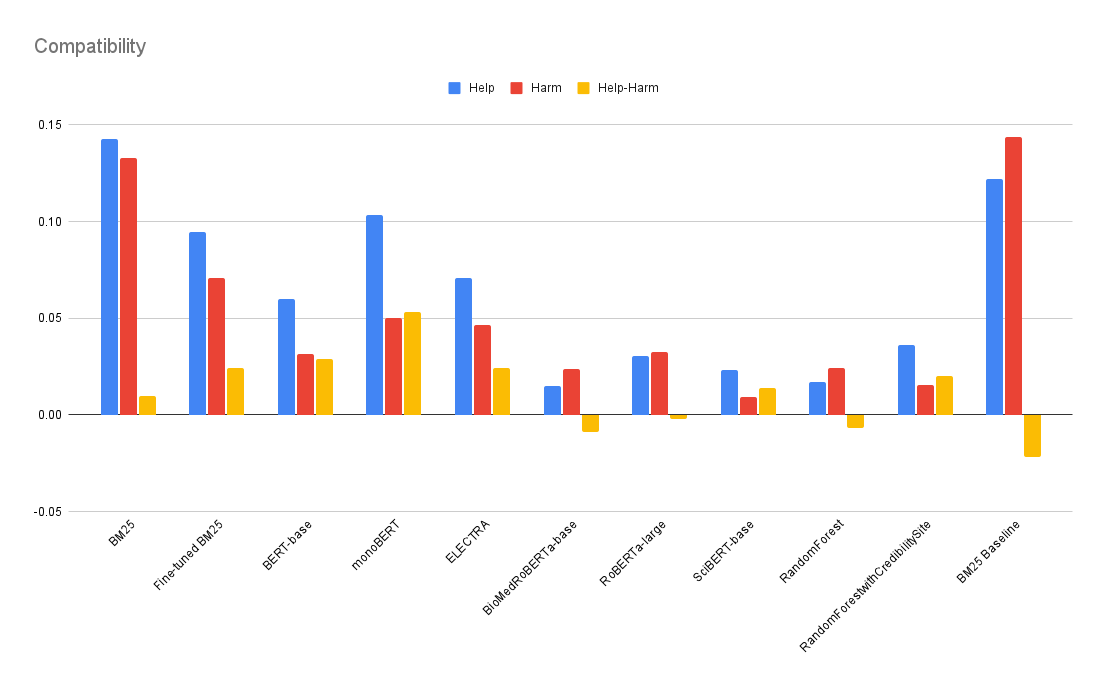}
    \caption{Individual Models Evaluated with Compatibility.}
    \label{imewc}
\end{figure*}

\end{document}